\documentclass{aa}  
\usepackage{graphicx}
\usepackage{natbib}
\usepackage{txfonts}
\bibpunct{(}{)}{;}{a}{}{,} 
 \def\mso{\,\mathrm{M}_\odot}
 
 \def\lso{\,{\rm L}_\odot}

 \def\llso{\log\, L/{\rm L}_\odot \,}
 \def\simle{\mathrel{\hbox{\rlap{\hbox{\lower4pt\hbox{$\sim$}}}\hbox{$<$}}}}
 \def\simgr{\mathrel{\hbox{\rlap{\hbox{\lower4pt\hbox{$\sim$}}}\hbox{$>$}}}}

\begin{document}

   \title{Blue supergiants as descendants of magnetic main sequence stars}

   \author{I. Petermann, N. Langer, N. Castro, \and L. Fossati
          }

   \institute{Argelander Institut f\"ur Astronomie der Universit\"at Bonn, Auf dem H\"ugel 71, 53121 Bonn, Germany
             }

   \date{Received date; accepted date}

  \abstract{
    About 10\% of the massive main sequence stars have recently been found to host 
    a strong, large scale magnetic field. Both, the origin and the evolutionary consequences of these
    fields are largely unknown.
    We argue that these fields may be sufficiently strong in the deep interior of the stars
    to suppress convection near the outer edge of their convective core. We performed parametrised
    stellar evolution calculations and assumed a reduced size of the convective core for stars 
    in the mass range 16$\mso$ to 28$\mso$ from the zero age main sequence until core carbon
    depletion.
   We find that such models avoid the coolest part of the main sequence band, which is usually filled
    by evolutionary models that include convective core overshooting. Furthermore, our `magnetic' models
    populate the blue supergiant region during core helium burning, i.e.,
    the post-main sequence gap left by ordinary single star models, and some of them
    end their life in a position near that of the progenitor of Supernova 1987A in the HR diagram. 
    Further effects include a strongly reduced luminosity during the red supergiant stage, and
    downward shift of the limiting initial mass for white dwarf and neutron star formation.
    }
%
   \maketitle
\section{Introduction}

Massive stars are important cosmic engines, because they drive the evolution of star forming galaxies (Mac Low et al. 2005),
and are very bright and thus visible from very long distances (Kudritzki et al. 2008).
Nevertheless, even their longest lasting evolutionary stage, core hydrogen burning,
has not yet been understood well. While one of the two culprits, mass loss, strongly affects only the very massive
stars, it is the other one, internal mixing processes, that still leads to many questions.   
This refers to thermally driven mixing as caused by convection, overshooting, or semiconvection as well
as to rotationally induced mixing and the mixing affected by magnetic fields.

In recent years, it has become evident that the concept of considering one evolutionary path for stars of a given
mass fails in the massive star regime. Hunter et al. (2008) and Brott et al. (2011) show that early B-type main sequence
stars fall into several distinctly different categories, as classified by their spin and surface nitrogen abundance.
Dufton et al. (2013) 
show that the rotational velocity distribution of LMC early B stars is bimodal, similar
to late B and A stars (Royer et al. 2007). The physical reasons for this are not clear yet,
but could relate to initial rotation, binarity, and strong internal magnetic fields (Langer 2012). 

In the present paper, we consider one of the least investigated aspects of massive star evolution,
namely the effects of internal magnetic fields. Four different types of magnetic fields in massive stars
have been discussed in recent years. First, there is the idea that differential rotation in their radiative
envelopes produces toroidal fields through a dynamo process (Spruit 2002). While the functioning of this dynamo is
still being debated (Zahn 2011), asteroseismic measurements in red giants (Mosser et al. 2012) call for a strong angular momentum
transport mechanism which could be provided by this predominantly toroidal field (Suijs et al. 2008; Heger et al. 2005).
This type of field, which would not be directly observable, and its consequences would be as ubiquitous 
as differential rotation. 

Second, massive stars have convective cores, and it appears likely that 
an $\alpha\Omega$-dynamo can produce a magnetic field in these cores. In fact, the MHD simulations
of Brun et al. (2005) show the functioning of such a dynamo in the convective cores of even slowly
rotating $2\mso$ stars, where the field growths saturate near their equipartition values.
MacGregor \& Cassinelli (2003) show that this field is unlikely to rise to the surface of the star
during its lifetime, so again, it may not be directly observable. Its effects in the vicinity of the 
convective core are largely unexplored.

A third type of magnetic field in massive stars has been suggested by Cantiello et al. (2009), who propose that
an $\alpha\Omega$-dynamo may work in their iron opacity driven sub-surface convection zones. The produced
flux tubes could then buoyantly float to the stellar surface and produce magnetic spots 
(Cantiello \& Braithwaite 2011), which may be observationally indicated by the moving discrete absorption
components in the UV resonance lines found in O-star spectra (Prinja \& Howarth 1988; Kaper et al. 1997). 

Here we are concerned with a fourth type of magnetic field in massive stars, namely
strong large scale fields that are anchored in the radiative stellar envelope and that form
an observable magnetosphere outside the star. Only OB~stars with such fields will be designated as
magnetic OB stars in this paper. Until recently, only three magnetic O-type stars were known, 
HD 37022 (Donati et al. 2002), HD 191612 (Donati et al. 2006) and $\xi$ Orionis\,A (Bouret et al. 2008). 
However, new intense searches have revealed many more magnetic OB~stars 
(Grunhut et al. 2009,  Petit et al. 2013, Hubrig et al. 2013, 2014), thus
implying an incidence fraction of strong, large scale B-fields in OB stars of about ten percent 
(Grunhut \& Wade 2012).

In contrast to the first three types of magnetic fields in OB stars, which all need a dynamo process
to continuously produce the field because it would quickly decay otherwise,
the strong large scale fields are thought to be stable for two reasons. First, they occur mostly in very 
slow rotators and in stars with radiative envelopes, such that there is no known dynamo process that could 
produce them. Second, Braithwate and Spruit (2004) find that magnetic field configurations
resembling the ones observed in magnetic OB stars can indeed be stable, in the sense that their decay
may take longer than the stellar lifetime. 
 
The origin of these stable, large-scale magnetic fields in massive stars is mainly described by two
ideas. The hypothesis of fossil fields (Braithwaite \& Nordlund 2006) assumes that magnetic fields in a molecular cloud
are conserved during star formation. It is, however, difficult to explain, why only a low
percentage of stars preserves a magnetic field in this formation channel. The effects of a fossil magnetic 
field on convective core dynamos in A-type stars are described in Featherstone et al. (2009).

Magnetic stars can also be the result of strong binary interaction (Langer 2014).
For example, they could be the product of the merger of two main sequence stars, which occurs 
quite frequently because a large number of massive stars have close companion stars
(Sana et al. 2012).
Garcia-Berro et al. (2012) show that strong magnetic fields can form 
in the hot, convective, differentially rotating envelope of a merger product of two white dwarfs,
which may occur similarly in the case of a main sequence merger.
Calculations by de Mink et al. (2014) suggest that
among the galactic early-type stars, 8$^{+9}_{-4}$ \% are the products of a merger event from a close binary system.
The order of magnitude obtained in this study agrees well with the occurrence of magnetic massive stars.
Furthermore, that many magnetic OB stars are nitrogen-enriched at their surface complies well with
the models of merger stars by Glebbeek et al. (2013).

The evolutionary consequences of strong large scale magnetic fields in the stellar interior are
largely unexplored. Most observed magnetic stars rotate slowly, such that rotational mixing
plays no role at least at the present time. Furthermore, it is found that in many situations, large scale
magnetic fields and convection repulse each other (e.g., Gough \& Tayler 1966; Galloway \& Weiss 1981).
Therefore, while in evolutionary models of massive stars it is usually assumed that the convective
core is extended over its standard size during core hydrogen burning
(Brott et al. 2011; Ekstr{\"o}m et al. 2012), we assume the opposite in the
present paper, which means that internal large scale magnetic fields restrict the convective core to a smaller
region than what the Ledoux criterion for convection would predict.
We elaborate our assumptions and the computational method in Sect.~2, and present our
evolutionary calculations in Sect.~3. After a comparison with observations in Sect.~4 we give
our conclusions in Sect.~5.

\section{Method}
Our stellar models are calculated with a hydrodynamic stellar evolution code, which is described 
in detail in Yoon et al. (2006) and K{\"o}hler et al. (2015).\\ 
Convection is modelled using the Ledoux criterion, assuming a mixing-length parameter $\alpha_{\rm{MLT}}$=1.5 
(Langer 1991). Semiconvection is implemented as described by Langer et al. (1983), applying
a value of $\alpha_{\rm{sem}}$ = 0.01 (Langer 1991; see also Zaussinger and Spruit 2013). 
The description for mass loss follows Vink et al. (2000)
for main sequence stars and blue supergiants, and Nieuwenhuijzen et al. (1990) for red supergiants.
For our models we adopt solar abundances as the initial composition as in Brott et al. (2011).

The observed fields in magnetic massive stars are large scale fields, and while some of them have a 
more complex geometry many have a dominant dipole component (Donati \& Landstreet 2009). Braithwaite \& Spruit (2004)
show that in order to achieve a stable magnetic field configuration, a dipole field
needs to be balanced by an internal toroidal field. While a typical value for the observable 
part of the dipole field is $1\,$kG (Fossati et al. 2015), magnetic flux conservation would then predict
field strength of up to $10^6\,$G near the convective core of a massive star. Additionally,
Braithwaite (2009) has shown that the toroidal component can have a much higher field strength
in a stable configuration than the poloidal component. Therefore,  
field strength of the order of $10^8\,$G cannot be excluded in the deep interior of magnetic massive stars and
this corresponds to the equipartition field strength and indicates direct dynamical effects
of the magnetic field on the convective motions.

Additionally, Zahn (2009) has shown that the magnetic field strength that is required to suppress convection is
about two orders of magnitude less than the equipartition field strength, when thermal and Ohmic
diffusion are taken into account. Zahn argues that this effect is observationally confirmed by the
presence of sunspots and by surface inhomogeneities in Ap/Bp stars. We therefore assume in the
following that convection near the boundary of the convective core is suppressed in magnetic massive
stars. 

Lydon \& Sofia (1995) have suggested a modification of the stellar structure equations
to account for the presence of a large scale magnetic field inside a star (see also
Feiden \& Chaboyer 2012). In our models, we ignore the changes induced by B fields
on the structure of the star, which implies that we assume that the magnetic energy density is lower than the thermal energy density.
We conduct a simple parametric study of stellar evolution models here with a diminished core mass
compared to what is predicted by the Ledoux criterion for convection. To this end,
we applied a modified Ledoux criterion as
\begin{equation}
\nabla_{rad} < f * \nabla_{Led}  , 
\end{equation} 
where
\begin{equation}
\nabla_{Led}=\nabla_{ad} - \frac{\chi_{\mu}}{\chi_{T}} \nabla_{\mu} .
\end{equation}
Whereas this mimics the convection criterion to some extent in the presence of a vertical
magnetic field proposed by Gough \& Tayler (1966; see also Schatten \& Sofia 1981; 
Tayler 1986), a quantitative prediction 
of the suppressive effects of large scale fields upon convection
requires multi-dimensional MHD calculations. We thus refrain from relating our parameter $f$
to magnetic field strengths, but make a heuristic 
choice for its values between 1.12 and 1.20, which leads to a
reduction of the convective core mass of the order of 
15\% in our models. 
We restricted this change to the core hydrogen-burning phase, because the interactions of the magnetic field with
the helium-burning core are uncertain. In particular, it remains unclear whether a significant fraction of the
magnetic field will survive below the hydrogen-burning shell source.

At present, the origin of the large scale fields in massive stars is not clear (Langer 2014).
While it might exist throughout core hydrogen burning if the field was inherited from 
the star formation process, it would be present for a fraction of that time if it was created
in a binary merger or mass transfer process. However, a significant mass gain is expected to 
rejuvenate the star, that is to say that the binary interaction resets its evolutionary clock. 
Even though the time is not reset to zero in this case, 
we assume that the effect of a magnetic field decreasing the convective core is unaltered
since the beginning of core hydrogen burning in our models. 

All our models are evolved through core carbon burning, such that their presupernova positions in 
the HR diagram can be reliably predicted.

\section{Results}
\label{results}
 \begin{table*}[tbp]
  \caption{Key quantities of the computed evolutionary models. Columns: acronym of the sequence,
   initial mass $M_{\rm i}$, core reduction parameter $f$, 
   the initial mass of the convective core $M_{\rm cc,i}$,
   initial and final helium core masses $M_{\rm He,i}$ and $M_{\rm He,f}$, lifetimes of core hydrogen 
   and core helium burning $\tau_{\rm H}$ and $\tau_{\rm He}$, 
   and presupernova effective temperature $T_{\rm eff,f}$, and luminosity $L_{\rm f}$}.
  \centering
  \begin{tabular}{ l l l | l l l l l l l }
     \hline\hline
  Acronym & $M_{\rm i}$ & $f$ & $M_{\rm cc,i}$ & $M_{\rm He,i}$ & $M_{\rm He,f}$ & $\tau_{\rm H}$ & $\tau_{\rm He}$ 
   & $T_{\rm eff,f}$ & $\log ( L_{\rm f}/L_{\odot} )$ \\
   ~ & $\mso$ & ~ & $\mso$ & $\mso$ & $\mso$ & Myr & Myr & kK & ~ \\
  \hline  
   M16f1.12 & 16 & 1.12 &  4.8 & 2.41    & 3.83 &     8.75 & 1.22  &   3.48  & 4.65 \\ 
   M16f1.16 & 16 & 1.16 &  4.4 & 2.27    & 3.77 &     8.56 & 1.36  &   3.51  & 4.61 \\
   M16f1.18 & 16 & 1.18 &  4.3 & 2.13    & 3.61 &     8.46 & 1.43  &   3.54  & 4.56 \\
   M16f1.20 & 16 & 1.20 &  4.1 & 2.12    & 3.52 &     8.37 & 1.37  &  17.4  & 4.70 \\
            &    &      &     &     &     &            &       &         &      \\
   M18f1.12 & 18 & 1.12 &  5.7 & 2.90    & 4.53 &     7.56 & 0.94  &  3.48   & 4.73 \\
   M18f1.16 & 18 & 1.16 &  5.2 & 2.68    & 4.44 &     7.38 & 1.04  &  3.49   & 4.72 \\
   M18f1.18 & 18 & 1.18 &  5.1 & 2.58    & 3.73 &     7.30 & 1.33  & 23.2   & 4.88 \\
   M18f1.20 & 18 & 1.20 &  5.0 & 2.40    & 3.32 &     7.21 & 1.46  & 20.5   & 4.85 \\
            &    &      &     &     &     &            &       &         &      \\
   M20f1.10 & 20 & 1.10 &  7.0 & 3.46    & 5.39 &     6.78  & 0.71  &  3.43  & 4.88 \\
   M20f1.13 & 20 & 1.13 &  6.5 & 3.28    & 4.33 &     6.66  & 0.92  & 22.9  & 5.01 \\
   M20f1.15 & 20 & 1.15 &  6.2 & 3.14    & 4.15 &     6.57  & 1.00  & 23.4  & 5.01 \\
            &    &      &     &     &     &             &       &        &      \\
   M22f1.18 & 22 & 1.18 &  7.0 &  3.36   & 4.30 &      5.82  & 0.93  & 24.6  & 5.10 \\
   M24f1.18 & 24 & 1.18 &  7.9 &  3.86   & 5.36 &      5.32  & 0.65  & 25.1  & 5.16 \\
   M28f1.18 & 28 & 1.18 & 10.0 &  4.76   & 6.54 &      4.61  & 0.50  & 24.7  & 5.31 \\
   \hline
\end{tabular}
\label{tab}
\end{table*}

\begin{figure}
  \centering
  \includegraphics[angle=0,width=8.95cm]{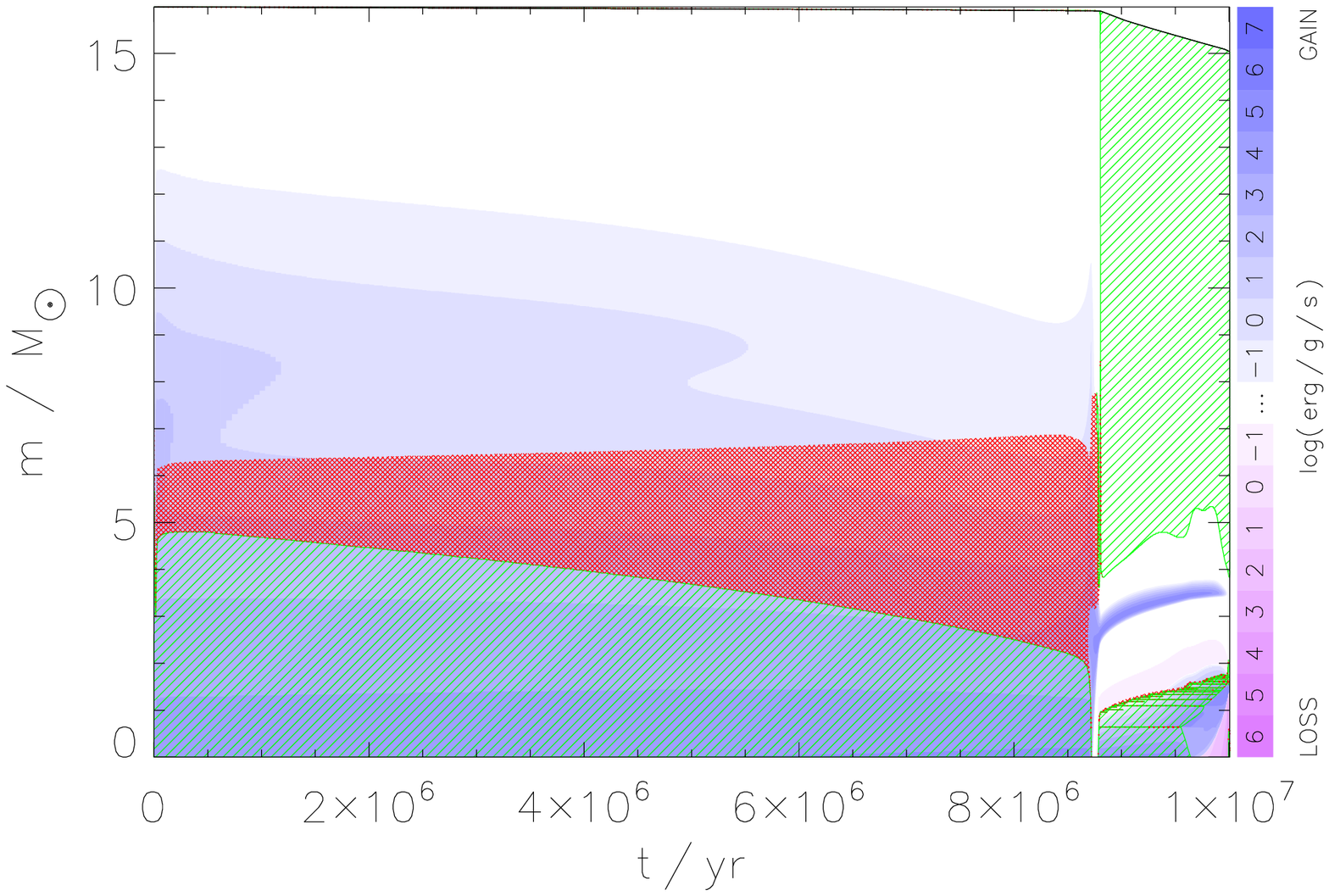}
  \includegraphics[angle=0,width=8.95cm]{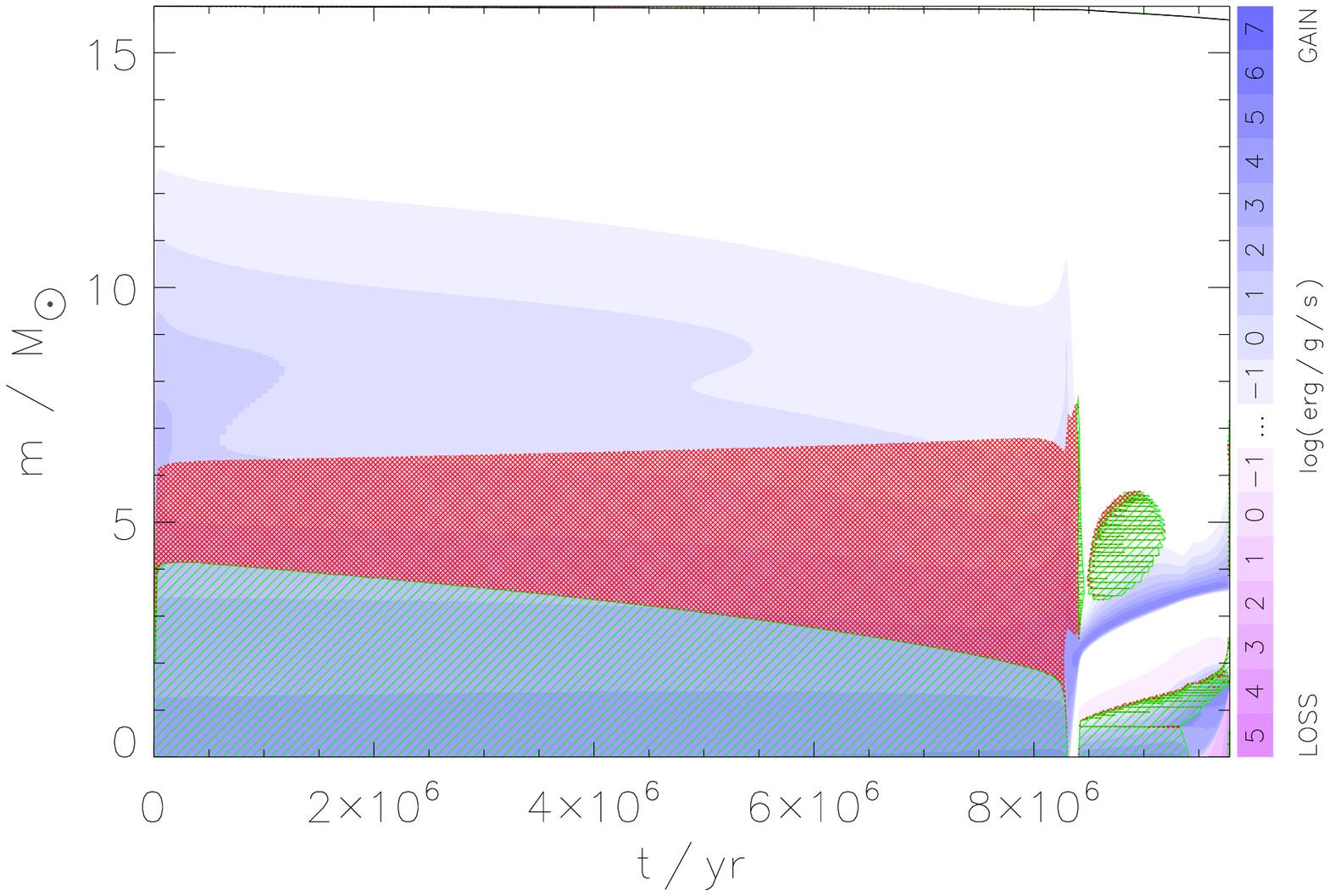}  
  \caption{Kippenhahn diagrams of two 16\,M$_{\odot}$ models computed with a convective core reduction parameter of $f$=1.12 (top)
         and $f$=1.20 (bottom). Convective (green) and semiconvective (red) 
         regions are shown as a function of mass coordinate and time. Thermonuclear energy generation or loss  is colour-coded as shown on 
         the scale on the right-hand side.
         The evolutionary tracks of both models in the HR diagram are shown in  Fig.~\ref{fig:161820M}}
 \label{fig:kipps}
\end{figure}

We performed calculations for stars in the mass range 16 to 28 M$_{\odot}$, assuming different 
values for the convective core reduction parameter $f$ (cf., Eq.~1), and Table\,\ref{tab} gives an overview of the
computed sequences.  
Figure~\ref{fig:kipps} provides Kippenhahn diagrams showing the internal evolution of two 16 M$_{\odot}$ star
as an example, where reduction parameters $f= 1.12$ (top) and $f= 1.20$ (bottom) have been adopted.
The convective cores are initially 5\ \% and 18\ \% lower in mass than for a non-reduced core,
which leads to initial helium core masses of 2.4$\mso$ and 2.1$\mso$, in both cases, compared to
2.9$\mso$ for the unreduced case.

As shown in Fig.~\ref{fig:kipps}, extended semiconvective shells develop on top of the convective cores
for the models with $f > 1$. While the interaction of these semiconvective layers with the large scale
magnetic field is unclear, it is likely that the field also inhibits any semiconvection. For our simulations, this would make little difference, since the chosen small
semiconvective efficiency parameter (see Sect.~2) means that there is very little chemical or energy transport 
in these zones.

The two stars shown in Fig.~\ref{fig:kipps} develop helium cores of initially  
about 2.6\,M$_{\odot}$ ($f= 1.12$) and 2.2\,M$_{\odot}$  ($f= 1.20$), respectively,
which increase to a final mass of about 3.5\,M$_{\odot}$ during core helium burning.
Such values are expected in stars in the mass range 11\,M$_{\odot}$ to 12\,M$_{\odot}$
for the physics used in the models of Brott et al. (2011), that is, relatively large 
convective core overshooting. 


%
Figure ~\ref{fig:161820M} (top) shows the evolutionary tracks of both 16\,M$_{\odot}$ models in the HR diagram. The minimum 
effective temperature during core hydrogen burning is higher for the model with the larger core reduction parameter,
but both are very high compared to that of a model by Brott et al. (2011) shown in the same plot. 
Further on, the tracks in Fig.~\ref{fig:161820M} show that the model computed with $f$=1.20 never evolves into
a red supergiant. Instead, its lowest effective temperature is about 10\,000\,K, and it ends its evolution
at about 17\,000\,K, which is close to the effective temperature of the progenitor of Supernova 1987A.
The 16\,M$_{\odot}$ model computed with $f$=1.12, on the other hand, moves to the red supergiant branch
quickly after core hydrogen exhaustion, and remains there for the rest of its evolution. Its luminosity
during core helium burning and its presupernova
luminosity of $\llso \simeq 4.65$ is much lower than the luminosity of models that assume convective core overshooting. 
\subsection{Variation in the core reduction parameter}
\begin{figure}[h!]
 \centering
  \includegraphics[angle=0,width=8.5cm]{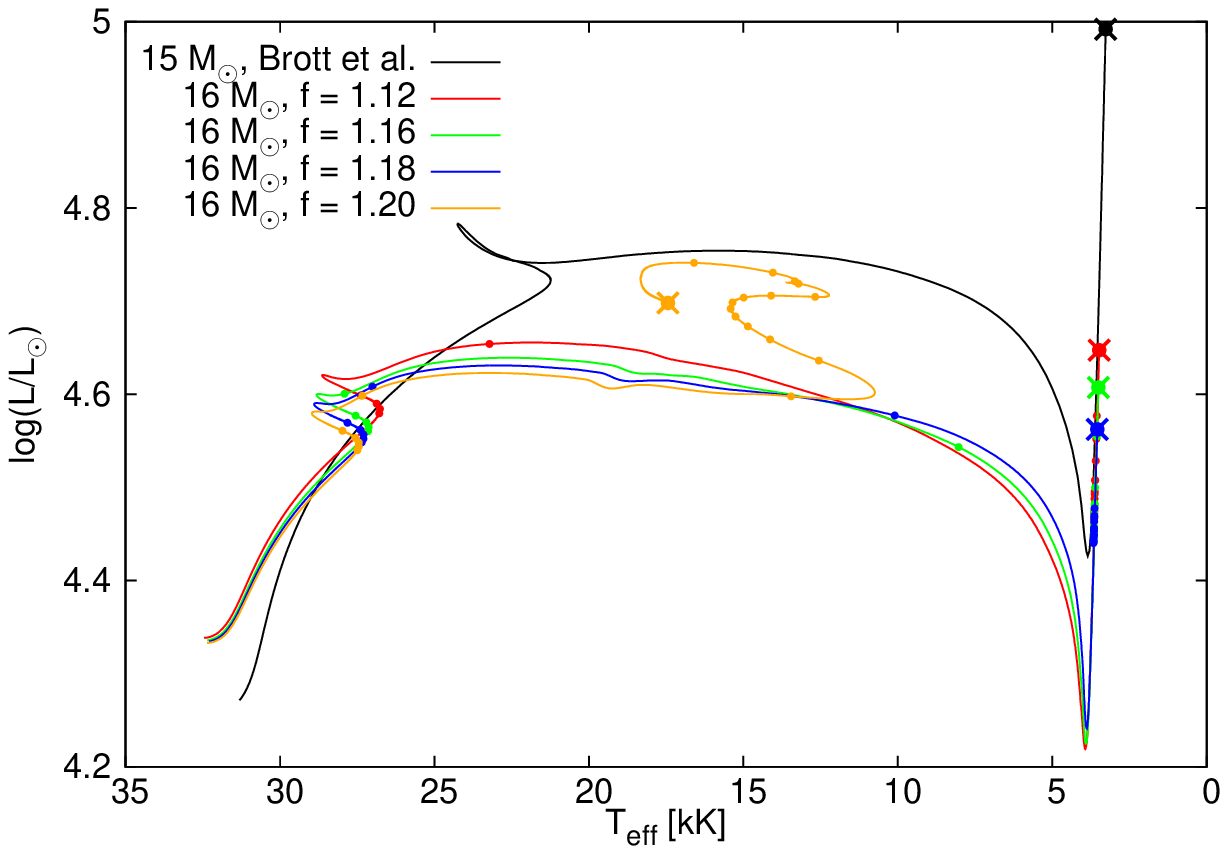} 
  \includegraphics[angle=0,width=8.5cm]{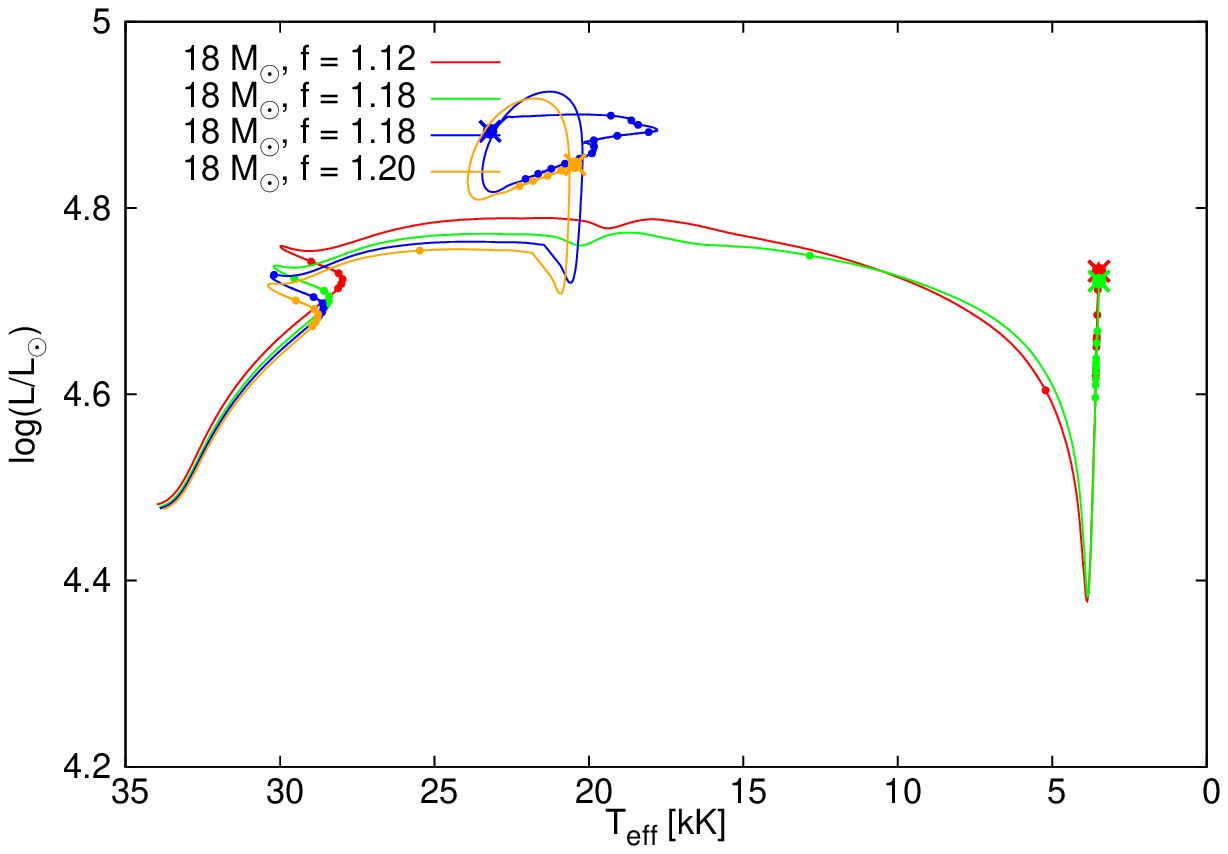}
  \includegraphics[angle=0,width=8.5cm]{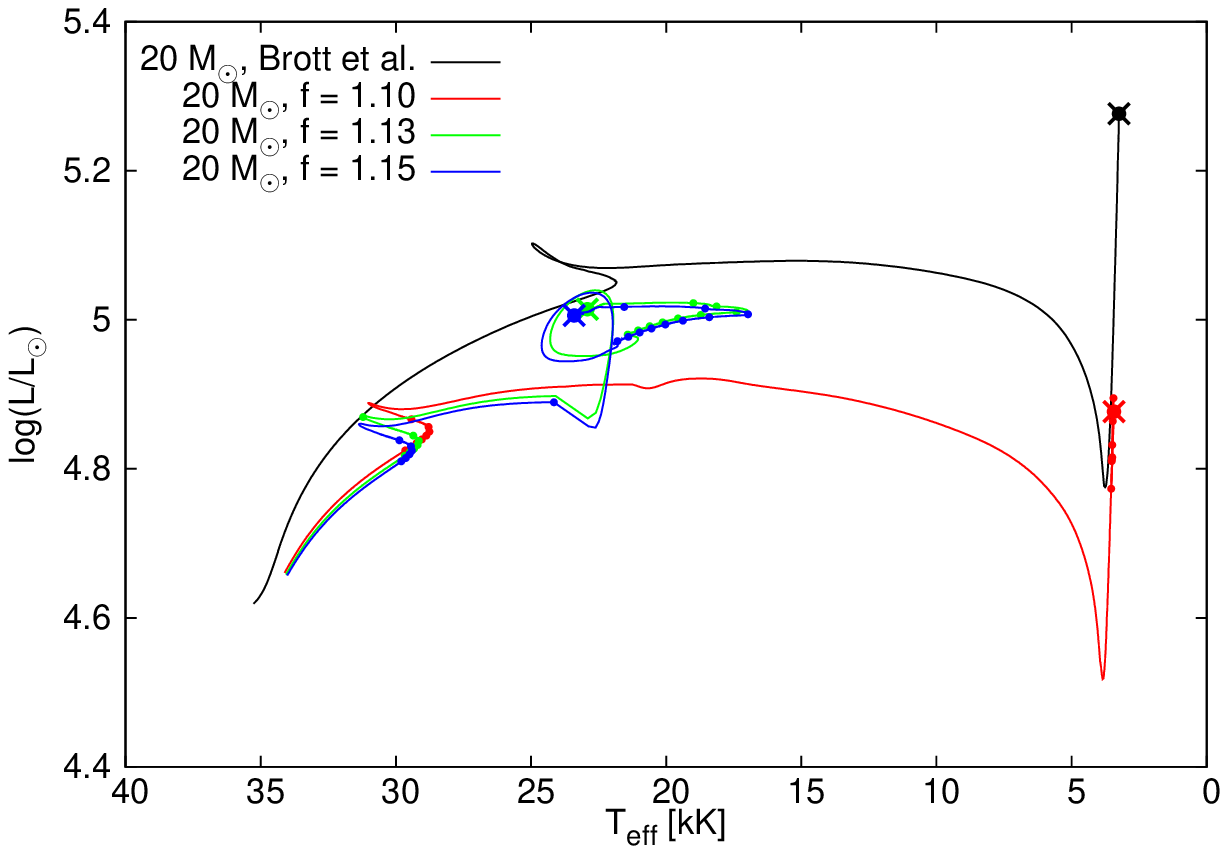}
  \caption{Evolutionary tracks of  stars with M=16, 18 and 20 M$_{\odot}$, computed with different
core reduction parameters as indicated, in the HR diagram. For the post main sequence phases, small dots
are placed on the tracks every 10$5$\,yr to indicate the speed of evolution of the models in the HR diagram.
   The large circle indicates the presupernova position of the models. 
  For comparison, tracks for 15 and 20\,M$_{\odot}$ stars of non-rotating models with solar metallicity by Brott et al. (2011) are also shown.}
 \label{fig:161820M}
\end{figure}

For stars of initially 16$\mso$, 18$\mso$, and 20$\mso$, we computed a series of models with different
reduction parameters $f$ (see Table \ref{tab}). The evolutionary tracks in the HR diagram of these models are shown in 
Fig.~\ref{fig:161820M}, together with tracks for non-rotating Galactic stars from Brott et al. (2011) for 15$\mso$ and 20$\mso$.

Figure~\ref{fig:161820M} reveals that the tracks in the HR diagram show discontinuous behaviour as a function
of the parameter~$f$. While at 16$\mso$, for example, the tracks vary little but in a systematic way 
for $f=1.12$, $1.16$, and $1.18$, with slightly reduced main sequence widths and presupernova 
red supergiant luminosities for higher values of $f$, the track computed with $f=1.20$ behaves qualitatively in a different way.
Its main sequence part is still within the trend of the tracks with lower $f$-values, but during core
helium burning, the model avoids the red supergiant regime, and it evolves into a presupernova configuration
that is significantly more luminous than for the three other 16$\mso$ sequences.  

Such discontinuous behaviour is also found at 18$\mso$ and 20$\mso$, although the jump occurs at different 
$f$-values. For the 18$\mso$ sequences, $f=1.16$ still leads to a red supergiant configuration,
whereas $f=1.18$ and $f=1.20$ does not. And in the 20$\mso$ models, the discontinuity is found between
$f=1.10$ and $f=1.13$. If $f$ were to measure the strength of the large scale magnetic field near
the convective core boundary of our models, less massive stars would require stronger fields in order
to undergo core helium burning as a blue rather than a red supergiant.

The discontinuous behaviour also concerns the presupernova position of the models in the HR diagram. 
The models that burn helium as blue supergiants also end their evolution in this regime, while
those that burn helium as red supergiants also die as such. Whether this remains true for intermediate 
values of $f$, which are not considered here, is not known. However, the 16$\mso$ model computed with
$f=0.20$ moves rather far towards the red supergiant regime after core hydrogen exhaustion and then
turns around, such that the model burns helium as a red supergiant. It might thus be possible that models
with $f$-values near the discontinuity have a brief red-supergiant phase before turning into core helium-burning
red supergiants. Similarly, all our core helium-burning blue supergiant models evolve redwards during 
core helium burning. However, those that are still blue supergiants at core helium exhaustion will 
produce blue presupernova stars, since the models evolve mostly bluewards during their post-core helium-
burning life time.  

Our `magnetic' red supergiant presupernova stars have remarkably low luminosities, compared to
stars computed with convective core overshooting, and their luminosity is smaller the larger
their core reduction parameter\,$f$. This can be seen from Fig.~\ref{fig:161820M}, which
shows the presupernova positions of our models in the HR diagram. For example, our
`magnetic' $16\mso$ models develop presupernova luminosities around 40\,000$\lso$,
whereas the $16\mso$ model of Brott et al. (2011) evolves into a presupernova red supergiant
with a luminosity of 100\,000$\lso$. From Fig.~\ref{fig:161820M} we conclude that 
our `magnetic' models would have to have a mass of about 22$\mso$ to produce such a luminous
red supergiant presupernova model.
 
As expected, models with larger $f$-parameters develop lower core masses (Table~\ref{tab}).  
In fact, the CO cores of our `magnetic' 16$\mso$ models barely exceed the Chandrasekhar mass.
This implies that, within our model assumptions, the limiting initial mass for neutron star
formation depends on the strength of the large scale B field. For the range of core reduction
parameters explored here, it would obviously not be found far below $16\mso$. A similar
shift may occur for the limiting initial mass for black hole formation.  

Figure\,\ref{fig:obs} shows the evolutionary tracks of models computed with a fixed value of the 
core reduction parameter\,$f$ in the HR diagram, considering the mass range $16\mso$ to $28\mso$.
Whereas the $16\mso$ model evolves into a red supergiant at the beginning of core helium burning, the higher
mass models all undergo core helium burning as blue supergiants. This confirms the trend that lower
mass models require larger core reduction parameters in order to remain on the blue side of the HR diagram. 
Furthermore, for this fixed value of~$f$, we find no significant trend in the effective temperature
as a function of mass for our `magnetic' core helium-burning blue supergiants. Instead, all the models are found
in the rather narrow effective temperature range of $T_{\rm eff}\simeq 17\dots23$\,kK.
While there seems to be no strong trend with mass, this does not imply that all `magnetic' models would
prefer this effective temperature range, as indicated, for example by our 16$\mso$ model computed with
$f=1.20$, which performs core helium burning at somewhat lower effective temperatures.

%

\subsection{Comparison to previous models}
\label{cm}
While we are not aware of massive star evolutionary models with reduced convective cores due to internal large
scale magnetic fields, models that arise from strong binary interactions can resemble 
our models to some extent. Considering the core helium-burning stage of evolution,
the main feature of our `magnetic' models is that the helium core mass is significantly lower 
than that of ordinary stars with the same initial mass. This effect is also produced in some
close binary models that simulate mass transfer, as well as in models for massive star mergers.

Braun \& Langer (1995) simulated binary mass transfer by considering single stars
that accreted considerable amounts of mass on a thermal time scale. They showed that, contrary
to models using the Schwarzschild criterion for convection, the finite mixing time
obtained when semiconvection is taken into account may prevent mass gainers from rejuvenating,
if the accretion occurs late enough during its core hydrogen-burning phase. In such models,
the convective core does not adapt to the new, higher mass, and thus the helium core mass 
in the later stages is significantly lower than that of ordinary stars. Braun \& Langer (1995)
found that for a given post accretion mass of 20$\mso$, a blue supergiant core helium-burning
evolution was obtained when the helium core mass was lower than a certain mass limit.
This is very analogous to the results we discuss in the previous section.

A similar result is obtained when mass transfer is initiated in a massive close binary just after
core hydrogen exhaustion (Claeys et al. 2011), or in a massive binary merger
that occurs after the more massive star has exhausted hydrogen in its core (Justham et al.
2014). For both situations, core helium-burning blue supergiant models are produced.
\section{Comparison with observations}
\subsection{Main sequence stars}
\begin{figure}[htbp]
  \centering
  \includegraphics[angle=0,width=8.5cm]{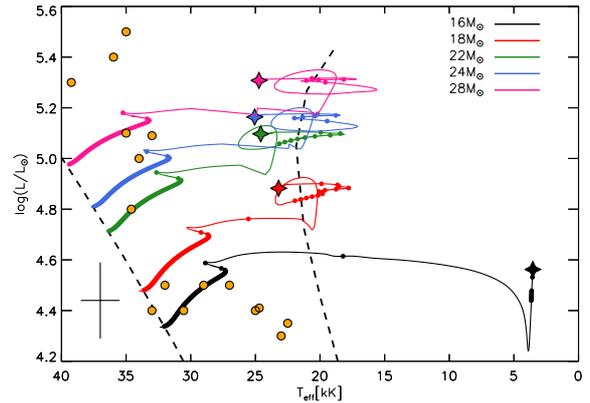}
  \caption{HR diagram for stars between 16 and 28 M$_{\odot}$ with convective cores reduced by the same factor, $f$=1.18. 
   Time steps for every 10$^5$ years are shown as small circles, starting at the end of the main sequence for clarity.
   The location of known massive magnetic main sequence stars (Briquet et al. 2013; Petit et al. 2013 and references
   therein; Alecian et al. 2014; Fossati et al. 2014; Neiner et al. 2014; Fossati et al. 2015; Castro et al. 2015)
   are shown as yellow dots. 
   Additionally, we draw the zero age and terminal age main sequences of the non-rotating models of Brott et al. (2011) as dashed lines.
            }
  \label{fig:obs}
\end{figure}
For the core hydrogen burning phase of evolution, the prime observable effect of our assumption 
that the large scale field leads to a lower convective core mass, is that the width of the main sequence
band in the HR diagram is reduced compared to ordinary models. This effect is clearly visible
in Fig.~\ref{fig:obs}, which shows the position of the terminal age main sequence (TAMS) of our
`magnetic' models and that of the non-rotating models of Brott et al. (2011). 

Figure~\ref{fig:obs} also shows the positions of the known massive magnetic main sequence stars in the
HR diagram. These stars are all Galactic objects. We find that the observed stars are not equally
distributed within the main sequence band of the Brott models, but that, with some exceptions, they indeed concentrate
in the area covered by the tracks of our `magnetic' models. We note that in a magnitude-limited
sample, which roughly holds for the observed magnetic stars, one instead expects the cool side of the
main sequence band to contain more stars than the hot side, since cool stars of the same luminosity
are visually brighter. This remains true even when considering that for a constant magnetic flux,
the magnetic field strength is less in more extended stars, and the field is thus harder to
detect (Fossati et al. 2015). Alternatively, the distribution of the magnetic stars in Fig.~\ref{fig:obs}
can also be explained by assuming that their B-field decays with a time scale close to or shorter than
their evolutionary time scale (see again Fossati et al. 2015). 

While this comparison does not give a final answer, it nevertheless shows that the assumption of 
magnetic stars having smaller convective cores than those of their non-magnetic counterparts
leads to a possible interpretation of the data.
In this context, it is interesting to mention that the convective core mass in massive stars may
be measured directly through asteroseismology, as suggested by Mazumdar \& Antia (2001)
and Mazumdar et al. (2006).
A comparative investigation of the convective core masses of magnetic and non-magnetic stars
through asteroseismology by Briquet et al. (2007, 2012) indeed supports the idea that the magnetic
stars do develop less massive cores.

If magnetic main sequence stars are really produced by the merging of two main sequence stars in a close
binary system (Langer 2014), and the large scale field leads to smaller convective cores, then it
is difficult to predict to what extent these objects will produce blue stragglers in star clusters
(Schneider et al. 2015).  
Such stars are more luminous and bluer than the bulk of the main sequence stars, meaning that they are located  
beyond the turnoff point in the HR-diagram of the cluster.
Even if the merged stars successfully rejuvenate, 
the subsequent suppression of the convective core mass may limit or even cancel the blue straggler effect;
that is to say the magnetic main sequence stars may or may not show up as blue stragglers.
On the other hand, blue stragglers may also form from slower accretion events, which are less likely
to give rise to large scale magnetic fields (Langer 2014). Thus, not all blue stragglers are
expected to be magnetic stars.

The observational evidence for magnetic fields in blue stragglers
is scarce, only about a dozen have been analysed so far. For roughly half of them magnetic fields were
unambiguously detected (see for example Bagnulo et al. 2006). Other high-precision analyses
could not confirm universal magnetic fields are present in blue stragglers (Fossati et al. 2010).

\subsection{Supergiants}

Figure~\ref{fig:obs} shows furthermore that our `magnetic' stars populate the region on the cool side of the
TAMS of the non-magnetic models of Brott et al. In fact, while observationally, this region appears to
be populated with stars (Fitzpatrick \& Garmany 1990; Hunter et al. 2008; Vink et al. 2010),
single star models generally do not spend any significant amount of time in this region (Ekstr{\"o}m et al. 2012;
Chieffi and Limongi 2013), such that the evolutionary history of the blue supergiants 
is in fact unclear at present (McEvoy et al. 2015).

Within the scope of the \textsc{VLT-FLAMES} Survey of Massive Stars, Evans et al. (2008), Hunter et al. (2008) and Vink et al. (2010)
all find a steep drop in rotation rates at $\log g \simeq 3.2$, as well as an increase in the surface nitrogen abundance. 
The quoted gravity defines the TAMS of the Brott et al. (2011) models around $15\mso$. Both features have been recently confirmed
within the \textsc{VLT-FLAMES} Tarantula Survey (Evans et al. 2011) by McEvoy et al. (2015).
Our `magnetic' models do provide an explanation for the observed population with $\log g < 3.2$. 

If the field in the magnetic main sequence stars does not decay,
our models would then predict that the blue supergiants with $\log g < 3.2$ should exhibit a large scale magnetic field.
Because their radii are larger by a factor of 3$\dots$10 and the main sequence stars show field strength of several hundred Gau{\ss}
(Petit et al. 2013; Fossati et al. 2015), flux conservation would lead to field strengths of the order of tens of Gau{\ss} 
or less (for these stars see Fig.~\ref{fig:obs}). These fields might thus be difficult to detect (Shultz et al. 2014).

As mentioned in Sect.~\ref{cm}, close binary evolution is also able to produce long-lived core helium-burning
blue supergiants just adjacent to the TAMS. It would thus be interesting to investigate the binary fraction of
the observed blue supergiants (Langer 2012). If they were produced by mass transfer or through a binary merger, detectable 
companion stars would hardly be expected (de Mink et al. 2014). The same would be true in the magnetic scenario
if the large scale fields were really produced by strong binary interaction (cf., Sect.~1).

Finally, as shown by our models (Fig.~\ref{fig:161820M}), some magnetic stars with smaller cores may also evolve into
red supergiants. We see two ways these might be observationally identified. One would be to look for the magnetic field itself.
While flux conservation would lead to inhibitively small average surface field strengths, convection in the deep convective envelope
might reorganise the field near the surface such that locally a much higher field strength could be possible. On the other
hand, the deep convective envelopes of red supergiants may initiate a dynamo process that could lead to fields in 
all red supergiants (Dorch 2004). Indeed, Grunhut et al. (2010) find B fields 
in a third of the 30 massive cool supergiants that they investigated.

The other characteristic of our `magnetic' red supergiant models is their low luminosity-to-mass ratio. This could show up in
the photometry of young star clusters showing red supergiants which are dimmer than expected from the main sequence turn-off.
Since the luminosity-to-mass ratio is thought to be proportional to the pulsation period in red supergiants
(Gough et al. 1965; Heger et al. 1997), objects with low luminosity-to-mass ratios might also
be identified through their pulsation properties.

\subsection{Supernovae}

When SN\,1987A exploded in the Large Magellanic Cloud, it was a big surprise that the progenitor star
turned out to be a blue supergiant (Walborn et al. 1989). While single-star models were
produced soon thereafter that did predict blue supergiant progenitors (e.g. Woosley 1988; 
Saio et al. 1988; Langer et al. 1989), single-star models that
include the OPAL opacities (Iglesias \& Rogers 1996) do not appear to produce blue supergiant
presupernova stars at the metallicity of the LMC.

As shown in Sect.\,\ref{results}, many of our `magnetic' models produce blue supergiant supernova progenitors.
On the other hand, as discussed in Sect.\,\ref{cm}, binary mass transfer and post main sequence merger may
also produce blue presupernova stars. In this context it is interesting to note that other SN\,1987A-like
supernovae have been observed (e.g., Kleiser et al. 2011, Pastorello et al. 2012), and in fact Smartt et al.
(2009) conclude that up to 3\% of all supernovae could be of this type, meaning they have a blue progenitor star.
Convective core suppression by large scale magnetic fields may help to explain such a large number 
of events.

\subsection{Magnetars}

Assuming flux freezing, the B field of the massive magnetic main sequence stars 
compressed to neutron star dimension would lead to the strength of magnetar fields
(Ferrario \& Wickramasinghe 2006). Indeed, if a massive main sequence stars had a B field 
of only $\sim 10^4\,$G in their core, the resulting B field in the neutron star would be of the order 
of $10^{14}\,$G,
which is two orders of magnitude greater than typical neutron star magnetic field strengths (Langer 2014).
In contrast to the scenario by Duncan \& Thompson
(1992), where the magnetar field forms from an extremely rapidly rotating collapsing iron
core, magnetars as successors of magnetic main sequence stars would form slowly rotating
neutron stars.

In the view of the idea that the magnetic fields in massive main sequence stars might be produced by
strong binary interaction,
Clark et al. (2014) propose that the blue hypergiant WD1-5 was the companion star
of the magnetar J1647-45 in the young Galactic cluster Westerlund\,1. To explain the unusual
properties of WD1-5, in particular its high helium and carbon surface abundances, they propose that
reverse mass transfer from the magnetar progenitor to WD1-5 must have occurred. This implies that
WD1-5 was the initially more massive star and the original mass donor in the binary system,
and that a strong accretion event could have formed a stable large scale field in the mass gainer.
Clearly, a suppression of the convective core mass in the magnetar progenitor would have helped to
obtain a supernova reversal, i.e. to have the initially less massive star explode first and produce
the magnetar in this event. 

\section{Discussion and conclusions}

We pursue the assumption that internal large scale magnetic fields in massive stars are capable of reducing the mass of
their convective cores during central hydrogen burning. While we are unable to 
quantitatively predict this effect, we take a parametric approach
by assuming that effectively larger temperature gradients are needed to produce convection compared to
the limit set by the Ledoux criterion (Gough \& Tayler 1986). 
Indeed, the ability of magnetic fields to restrict convection is documented well for low-mass stars.
The finding of Jackson et al. (2009) that the radii for low-mass M dwarfs in the young open cluster NGC\,2516 
are up to 50\% larger than expected is reconciled well by MacDonald \& Mullan (2013),
who apply the Gough \& Tayler criterion.
Feiden \& Chaboyer (2013, 2014), who model magnetic low-mass stars based on the technique developed by
Lydon \& Sofia (1985), confirm very similar behaviour by comparing them to detached eclipsing
binaries.

We followed the evolution of stars in the mass range 16-28 M$_{\odot}$
from the zero age main sequence roughly until neon ignition. 
Our ansatz allowed us to smoothly vary the extent of the core mass reduction and study its effects on the
star's main sequence and post-main sequence evolution. 

We showed that the stellar models obtained in this way predict various observational consequences.
Their lower core masses compared to ordinary models lead to an earlier termination of the main
sequence evolution, and to blue supergiant core helium-burning stars adjacent to the terminal age main 
sequence of ordinary stars. Both effects appear to be supported by observations, because the magnetic 
massive main sequence stars appear to prefer the hot side of the main sequence band, and
the observed hot blue supergiants are not predicted by single-star evolution otherwise.  
Our models may also help to explain SN1987A-like supernovae, perhaps more abundantly
than binary evolution.

On the other hand, binary evolution per se could produce the phenomenon of large scale
stable B fields in a fraction of the massive main sequence stars, such as through stellar mergers. 
Multidimensional MHD calculations are required to investigate this question, and
binary evolution models could then assess the effects of B fields on the further evolution
in more detail.

\begin{acknowledgements}
We are grateful to Stephen Smartt and Henk Spruit for enlightening discussions.
IP acknowledges support provided by DFG-grant LA 587/18.
LF acknowledges financial support from the Alexander von Humboldt Foundation.
\end{acknowledgements}
\section{References}
Alecian, E., Kochukhov, O., Petit, V., et al. 2014, A\&A, 567, A28\\
Bagnulo S., Landstreet J.D., Mason E., et al., 2006, A\&A, 450, 777\\
Bouret J.C., Donati J.F., Martins F., et al., 2008, MNRAS, 389, 75\\
Braithwaite J., 2009, MNRAS, 397, 763\\
Braithwaite J., Nordlund \r{A}., 2006, A\&A, 450, 1077\\
Braithwaite J., Spruit H.C., 2004, Nature, 431, 819\\
Braun H., Langer N., 1995, A\&A, 297, 483\\
Briquet M., Morel T., Thoul A., et al., 2007, MNRAS, 381, 1482\\
Briquet M., Neiner C., Aerts C., et al., 2012, MNRAS, 427, 483\\
Briquet, M., Neiner, C., Leroy, B., P\'{a}pics, P. I., \& MiMeS Collaboration. 2013, A\&A, 557, L16\\
Brott I., de Mink S.E., Cantiello M., et al., 2011, A\&A, 530, A115\\
Brun A.S., Browning M.K., Toomre J., 2005, ApJ, 629, 461\\
Cantiello M., Braithwaite J., 2011, A\&A, 534, A140\\
Cantiello M., Langer N., Brott I., et al., 2009, A\&A, 499, 279\\
Castro N., et al. 2015, submitted\\
Chieffi A., Limongi M., 2013, ApJ, 764, 21\\
Claeys J.S.W., de Mink S.E., Pols O.R., Eldridge J.J., Baes M., 2011, A\&A, 528, A131\\
Clark J.S., Ritchie B.W., Najarro F., Langer N., Negueruela I., 2014, A\&A, 565, A90\\
de Mink S.E., Sana H., Langer N., Izzard R.G., Schneider F.R.N., 2014, ApJ, 782, 7\\
Donati J.F., Babel J., Harries T.J., et al., 2002, MNRAS, 333, 55\\
Donati J.F., Howarth I.D., Bouret J.C., et al., 2006, MNRAS, 365, L6\\
Donati, J.-F. \& Landstreet, J. D. 2009, ARA\&A, 47, 333 \\ 
Dorch S.B.F., 2004, A\&A, 423, 1101\\
Dufton P.L., Langer N., Dunstall P.R., et al., 2013, A\&A, 550, A109\\
Duncan R.C., Thompson C., 1992, ApJ, 392, L9\\
Ekstr{\"o}m S., Georgy C., Eggenberger P., et al., 2012, A\&A, 537, A146\\
Evans C., Hunter I., Smartt S., et al., 2008, The Messenger, 131, 25\\
Evans C.J., Taylor W.D., H\'{e}nault-Brunet V., et al., 2011, A\&A, 530, A108\\
Featherstone, N. A., Browning, M. K., Brun, A. S., \& Toomre, J. 2009, ApJ, 705, 1000\\
Feiden, G. A. \& Chaboyer, B. 2012, ApJ, 761, 30\\
Ferrario L., Wickramasinghe D., 2006, MNRAS, 367, 1323\\
Fitzpatrick E.L., Garmany C.D., 1990, ApJ, 363, 119\\
Fossati L., Mochnacki S., Landstreet J., Weiss W., 2010, A\&A, 510, A8\\
Fossati L., Castro N., Morel T., et al., 2015, A\&A, 574, A20\\
Fossati, L., Zwintz, K., Castro, N., et al. 2014, A\&A, 562, A143\\
Galloway D.J., Weiss N.O., 1981, ApJ, 243, 945\\
Garc\'{i}a-Berro, E., Lor\'{e}n-Aguilar, P., Aznar-Sigu\'{a}n, G., et al. 2012, ApJ, 749, 25\\
Glebbeek E., Gaburov E., Portegies Zwart S., Pols O.R., 2013, MNRAS, 434, 3497\\
Gough D.O., Tayler R.J., 1966, MNRAS, 133, 85\\
Gough D.O., Ostriker J.P., Stobie R.S., 1965, ApJ, 142, 1649\\
Grunhut J.H., Wade G.A., Marcolino W.L.F., et al., 2009, MNRAS, 400, L94\\
Grunhut J.H., Wade G.A., Hanes D.A., Alecian E., 2010, MNRAS, 408, 2290\\
Grunhut J.H., Wade G.A., MiMeS Collaboration, 2012, In: Hoffman J.L., Bjorkman J., Whitney B. (eds.) American Institute of Physics Conference
Series, vol. 1429 of American Institute of Physics Conference Series, 67-74\\
Heger A., Jeannin L., Langer N., Baraffe I., 1997, A\&A, 327, 224\\
Heger, A., Woosley, S. E., \& Spruit, H. C. 2005, ApJ, 626, 350\\
Hubrig S., Sch{\"o}ller M., Ilyin I., et al., 2013, A\&A, 551, A33\\
Hubrig S., Fossati L., Carroll T.A., et al., 2014, A\&A, 564, L10\\
Hunter I., Lennon D.J., Dufton P.L., et al., 2008, A\&A, 479, 541\\
Iglesias C.A., Rogers F.J., 1996, ApJ, 464, 943\\
Jackson, R. J., Jeffries, R. D., \& Maxted, P. F. L. 2009, MNRAS, 399, L89\\
Justham, S., Podsiadlowski, P., \& Vink, J. S. 2014, ApJ, 796, 121\\
Kaper L., Henrichs H.F., Fullerton A.W., et al., 1997, A\&A, 327, 281\\
Kleiser I.K.W., Poznanski D., Kasen D., et al., 2011, MNRAS, 415, 372\\
K{\"o}hler K., Langer N., de Koter A., et al., 2015, A\&A, 573, A71\\
Kudritzki R.P., Urbaneja M.A., Bresolin F., et al., 2008, ApJ, 681, 269\\
Langer N., 1991, A\&A, 252, 669\\
Langer N., 2012, ARA\&A, 50, 107\\
Langer N., 2014, In: IAU Symposium, vol. 302 of IAU Symposium, 1-9\\
Langer N., Fricke K.J., Sugimoto D., 1983, A\&A, 126, 207\\
Langer N., El Eid M.F., Baraffe I., 1989, A\&A, 224, L17\\
Lydon, T. J. \& Sofia, S. 1995, ApJS, 101, 357\\
Mac Low M.M., Balsara D.S., Kim J., de Avillez M.A., 2005, ApJ, 626, 864\\
MacDonald, J. \& Mullan, D. J. 2013, ApJ, 765, 126\\
MacGregor K.B., Cassinelli J.P., 2003, ApJ, 586, 480\\
Mazumdar, A. \& Antia, H. M. 2001, A\&A, 377, 192\\
Mazumdar, A., Briquet, M., Desmet, M., \& Aerts, C. 2006, A\&A, 459, 589\\
McEvoy C.M., Dufton P.L., Evans C.J., et al., 2015, A\&A, 575, A70\\
Mosser, B., Goupil, M. J., Belkacem, K., et al. 2012, A\&A, 548, A10\\
Neiner, C., Tkachenko, A., \& MiMeS Collaboration. 2014, A\&A, 563, L7\\
Nieuwenhuijzen, H. \& de Jager, C. 1990, A\&A, 231, 134\\
Pastorello A., Pumo M.L., Navasardyan H., et al., 2012, A\&A, 537, A141\\
Petit V., Owocki S.P., Wade G.A., et al., 2013, MNRAS, 429, 398\\
Prinja R.K., Howarth I.D., 1988, MNRAS, 233, 123\\
Royer, F., Zorec, J., \& G{'o}mez, A. E. 2007, A\&A, 463, 671\\
Saio H., Nomoto K., Kato M., 1988, Nature, 334, 508\\
Sana H., de Mink S.E., de Koter A., et al., 2012, Science, 337, 444\\
Schatten, K. H. \& Sofia, S. 1981, Astrophys. Lett., 21, 93\\
Schneider, F. R. N., Izzard, R. G., Langer, N., \& de Mink, S. E. 2015, ApJ, 805, 20\\
Shultz M., Wade G.A., Petit V., et al., 2014, MNRAS, 438, 1114\\
Smartt S.J., Eldridge J.J., Crockett R.M., Maund J.R., 2009, MNRAS, 395, 1409\\
Spruit, H. C. 2002, A\&A, 381, 923\\
Suijs, M. P. L., Langer, N., Poelarends, A.-J., et al. 2008, A\&A, 481, L87\\
Tayler, R. J. 1986, MNRAS, 220, 793\\
Vink J.S., Brott I., Gr{\"a}fener G., et al., 2010, A\&A, 512, L7\\
Vink, J. S., de Koter, A., \& Lamers, H. J. G. L. M. 2000, A\&A, 362, 295\\
Walborn N.R., Prevot M.L., Prevot L., et al., 1989, A\&A, 219, 229\\
Woosley S.E., 1988, ApJ, 330, 218\\
Yoon, S.-C., Langer, N., \& Norman, C. 2006, A\&A, 460, 199\\
Zahn J.P., 2009, Communications in Asteroseismology, 158, 27\\
Zahn J.P., 2011, In: Neiner C., Wade G., Meynet G., Peters G. (eds.) IAU Symposium, vol. 272 of IAU Symposium, 14-25\\
Zaussinger F., Spruit H.C., 2013, A\&A, 554, A119\\
\end{document}